\PassOptionsToPackage{table}{xcolor} 

\documentclass[sigconf]{acmart}

\AtBeginDocument{%
  }

\setcopyright{acmlicensed}
\copyrightyear{2025}
\acmYear{2025}
\acmDOI{XXXXXXX.XXXXXXX}

\acmConference[Preprint]{Make sure to enter the correct
  conference title from your rights confirmation email}{2025
  }

\acmISBN{978-1-4503-XXXX-X/2018/06}

\usepackage{enumitem}
\usepackage{multirow}
\renewcommand\footnotetextcopyrightpermission[1]{}

\begin{document}

\title{VLM-based Prompts as the Optimal Assistant for Unpaired Histopathology Virtual Staining}

\author{Zizhi Chen}
\authornote{Equal contribution.}
\affiliation{%
  \institution{Fudan University}
  \country{China}
}
\email{chenzz24@m.fudan.edu.cn}

\author{Xinyu Zhang}
\authornotemark[1]
\affiliation{%
  \institution{Central South University}
  \country{China}
}
\email{8208220519@csu.edu.cn}

\author{Minghao Han}
\authornotemark[1]
\affiliation{%
  \institution{Fudan University}
  \country{China}
}
\email{mhhan22@m.fudan.edu.cn}

\author{Yizhou Liu}
\affiliation{%
  \institution{Harbin Institute of Technology}
  \country{China}
}
\email{kevinclaint.liu@gmail.com}

\author{Ziyun Qian}
\affiliation{%
  \institution{Fudan University}
  \country{China}
}
\email{zyqian22@m.fudan.edu.cn}

\author{Weifeng Zhang}
\affiliation{%
  \institution{Central South University}
  \country{China}
}
\email{8203211723@csu.edu.cn}

\author{Xukun Zhang}
\affiliation{%
  \institution{Fudan University}
  \country{China}
}
\email{zhangxk21@m.fudan.edu.cn}

\author{Jingwei Wei}
\authornote{Corresponding author}
\affiliation{%
  \institution{Chinese Academy of Sciences}
  \country{China}
}
\email{weijingwei2014@ia.ac.cn}

\author{Lihua Zhang}
\authornotemark[2]
\affiliation{%
  \institution{Fudan University}
  \country{China}
}
\email{lihuazhang@fudan.edu.cn}

\renewcommand{\shortauthors}{Zizhi Chen, Xinyu Zhang and Minghao Han, \textit{et al.}}

\begin{abstract}
 In histopathology, tissue sections are typically stained using common H\&E staining or special stains (MAS, PAS, PASM, \emph{etc.}) to clearly visualize specific tissue structures. The rapid advancement of deep learning offers an effective solution for generating virtually stained images, significantly reducing the time and labor costs associated with traditional histochemical staining. However, a new challenge arises in separating the fundamental visual characteristics of tissue sections from the visual differences induced by staining agents. Additionally, virtual staining often overlooks essential pathological knowledge and the physical properties of staining, resulting in only style-level transfer. To address these issues, we introduce, for the first time in virtual staining tasks, a pathological vision-language large model (VLM) as an auxiliary tool. We integrate contrastive learnable prompts, foundational concept anchors for tissue sections, and staining-specific concept anchors to leverage the extensive knowledge of the pathological VLM. This approach is designed to describe, frame, and enhance the direction of virtual staining. Furthermore, we have developed a data augmentation method based on the constraints of the VLM. This method utilizes the VLM’s powerful image interpretation capabilities to further integrate image style and structural information, proving beneficial in high-precision pathological diagnostics. Extensive evaluations on publicly available multi-domain unpaired staining datasets demonstrate that our method can generate highly realistic images and enhance the accuracy of downstream tasks, such as glomerular detection and segmentation. Our code is available at: \url{https://github.com/CZZZZZZZZZZZZZZZZZ/VPGAN-HARBOR} 
\end{abstract}





\maketitle

\begin{figure}[tp]
 \includegraphics[width=0.95\linewidth]{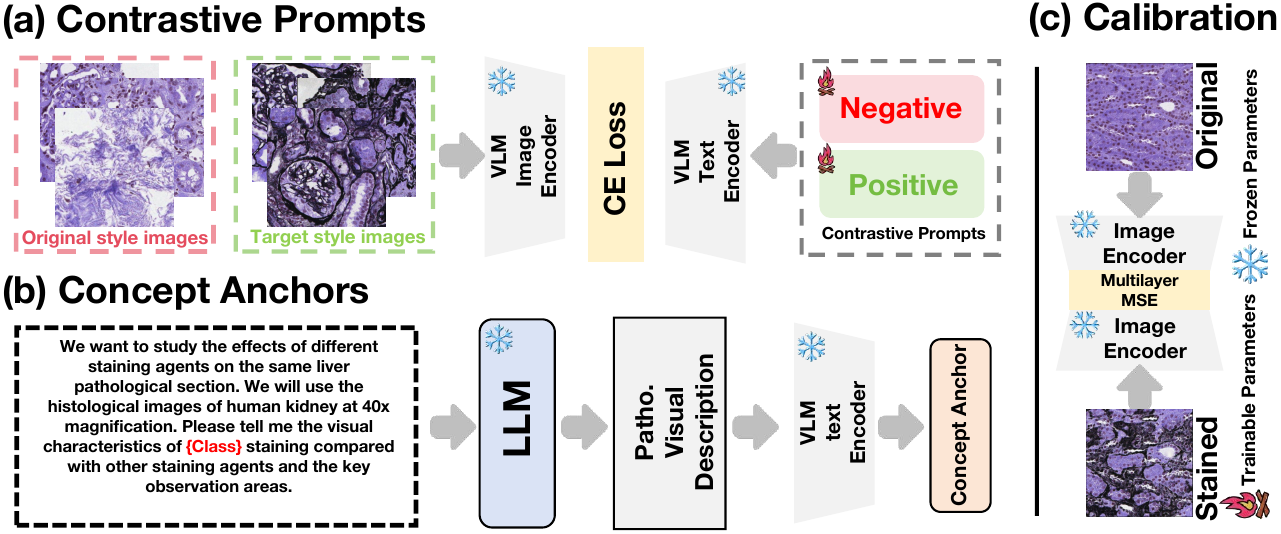}
\caption{Three auxiliary methods for virtual staining tasks proposed by us using the VLM: (a) Learnable contrastive prompts based on the classification task. (b) Concept anchors design based on LLM. (c) Visual calibration based on the VLM.}
\label{fig1}
\end{figure}

\section{Introduction}
Histopathological examination is widely regarded as the gold standard for the clinical diagnosis of diseases~\cite{CLAM,UNI,ABMIL}. This process typically involves histochemical staining, which differentiates various tissue components through distinct colors to aid in pathological diagnosis. Routine pathological examination commonly employs Hematoxylin and Eosin (H\&E) staining to highlight tissue morphology for initial diagnostic purposes. However, H\&E staining often fails to provide sufficient diagnostic information for many diseases. Consequently, the use of special stains~\cite{sstain1,sstain2} offers critical diagnostic insights across multiple dimensions~\cite{UMDST}. For instance, in renal pathology, Masson's Trichrome (MAS) staining is utilized to distinguish collagen fibers from muscle fibers, Periodic Acid-Schiff (PAS) staining is employed to better visualize the Glomerular basement membrane, Tubular basement membrane, and Mesangial matrix (GTM)~\cite{GTM1,GTM2}, and Periodic Acid-Schiff with Methenamine Silver (PASM) staining can more clearly delineate the GTM on the basis of PAS~\cite{PASM1,PASM2}. Nevertheless, the application of special stains generally requires more time and incurs higher labor costs. Moreover, when patients suffer from non-neoplastic kidney diseases~\cite{nonneoplastickidneydisease}, liver cirrhosis~\cite{cirrhosisdisease}, or other conditions, pathologists may necessitate multiple types of special stains to achieve a more accurate diagnosis. Therefore, the development of virtual staining technology, which reduces the costs of special stains for both pathologists and patients while addressing the need for multi-staining on the same tissue section, holds significant importance in clinical practice~\cite{cost1,cost2}.

Recent advancements in generative model~\cite{GAN,DDPM,SDEdiffusion} technology have also spurred progress in the image-to-image (I2I) domain within pathology~\cite{BCI,APT,UMDST,KIN}. These advances enable stain style transfer for color normalization~\cite{CAGAN,TTSaD} and serve as a feature enhancement strategy~\cite{PathUP}. They also train effective feature extractors, boosting performance in subtype classification~\cite{DiffusionMIL,pathoduet}. However, current pathological I2I methods largely follow the technical frameworks established in the natural image domain, focusing primarily on style transformation while neglecting the texture and cytological structures of pathological sections, as well as the physical and chemical properties of staining agents. This limitation undermines the reliability of virtually stained sections as diagnostic tools. In our view, expecting current generative models to possess the extensive domain knowledge and cellular-level visual discernment required in pathology is overly demanding. Such models urgently require a pathology expert-level "assistant" to aid them in more effectively accomplishing virtual staining tasks.

The advent of the VLM in pathology~\cite{plip,CONCH,MUSK,quilt1M,PathGen1.6M,pathasst,cpathomini} has made this endeavor feasible. Empowered by millions of pathology image-caption pairs, it possesses extensive pathological knowledge and robust capabilities in pathological image recognition. It has achieved state-of-the-art performance in various tasks, including pathology image classification, segmentation, caption generation, text-to-image synthesis, image and text retrieval. Given its role as an expert-level assistant in clinical decision-making, providing comprehensive support to pathologists, we aspire to extend its all-round excellence to the field of virtual staining. The provision of advanced pathological knowledge by the VLM is expected to potentially ensure that virtual staining results meet medical and chemical standards. Furthermore, intermediate staining processes could be characterized and fine-grained visual details might be captured through the leveraging of the VLM’s powerful multimodal information extraction capability, potentially leading to enhanced performance of virtual staining.

In this paper, we present three attempts to leverage VLMs for guiding virtual staining tasks. As shown in Figure \ref{fig1}(a), the \textbf{contrastive prompting tuning} employs contrastive learning strategies and binary classification tasks to decode and extract the rich information embedded in pathology VLMs. This enables the system to articulate stain differentiation and staining processes that are typically challenging to describe in human language. The \textbf{conceptual anchoring method} as presented in Figure \ref{fig1}(b) generates foundational and stain-specific concept anchors by leveraging the rich corpora produced by Large Language Model (LLM)~\cite{GPT4,deepseekR1} and the information compression capabilities of pathology VLM, guiding the "variation and invariance" during the staining process. For these two prompting strategies, we designed the \textbf{C}ontrastive \textbf{P}rompt \textbf{T}ransfer (CPT), \textbf{C}onstant \textbf{C}oncept \textbf{A}nchoring (CCA), and \textbf{I}ndependent \textbf{C}oncept \textbf{R}einforcement (ICR) modules, respectively. Together with a unpaired I2I model as baseline, these components form our method, the \textbf{V}LM-based \textbf{P}rompts \textbf{G}enerative \textbf{A}dversarial \textbf{N}etwork (VPGAN), which, to the best of our knowledge, represents the first attempt to bridge GANs and pathology VLM. Additionally, inspired by Xiong \emph{et al.}~\cite{DPI}, we argue that inference enhancement based on DDIM~\cite{DDIM} can effectively meet the demands of high-resolution diagnostic tasks. However, in practice, existing methods risk visual domain collapse (\emph{e.g.} H\&E2PASM). To address this, as illustrated in Figure ~\ref{fig1}(c), we adopted a \textbf{multi-level calibration strategy} based on the VLM, successfully improving the stability and performance of inference enhancement. Built on CLIP~\cite{CLIP} with ResNet101~\cite{resnet}, our method dynamically adjusts texture and color details across network layers, significantly improving inference robustness. The inference enhancement method empowered by VPGAN and the multi-level calibration strategy together constitute our \textbf{H}istopathology st\textbf{A}ining expe\textbf{R}t \textbf{B}ased \textbf{O}n p\textbf{R}ompts (HARBOR). The contribution of this paper is summarized as follows:

\begingroup
\setlist[itemize]{noitemsep, topsep=0pt, left=10pt}
\begin{itemize}
    \item As far as we know, our proposed VPGAN is the first GAN based on diversified prompts from the pathological VLM, which serves as a super assistant for virtual staining.
    \item We designed a VLM-based multi-level visual calibration module to tackle data staining domain disintegration and enhance data augmentation stability and performance.
    \item Our proposed method produced satisfactory images in three virtual staining tasks and showed optimal performance under different inference strategies, indicating it can meet diverse scenario, cost and detection task requirements.  
    \item Segmentation and detection accuracy across diverse glomerulus datasets are improved by our method, and strong clinical potential is demonstrated.
\end{itemize}
\endgroup

\begin{figure*}[tp]
 \includegraphics[width=0.9\linewidth]{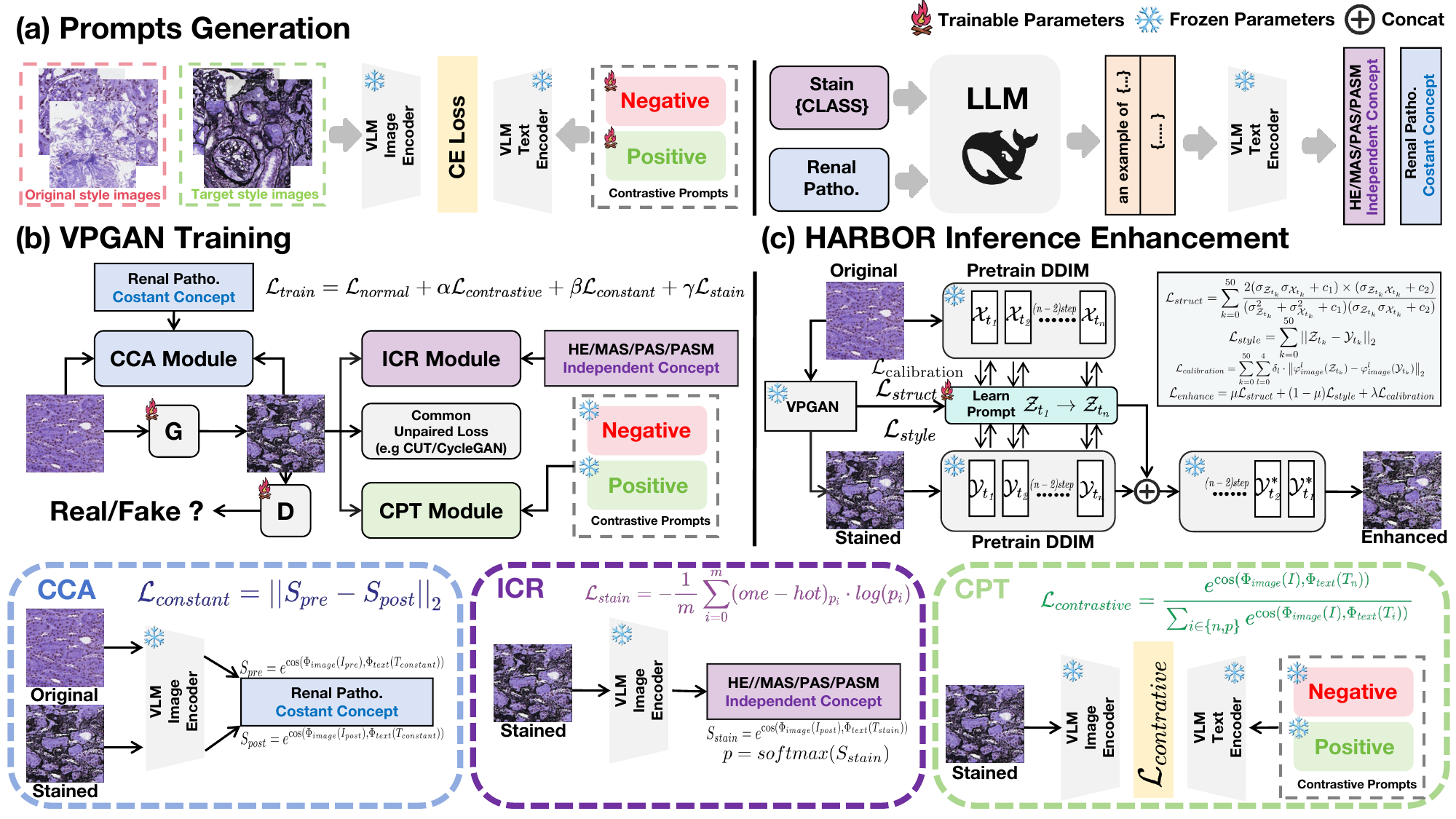}
\caption{Overview of the proposed VPGAN and HARBOR. In the prompt generation phase, we employed prompt tuning method to generate contrastive prompts based on a binary classification task. Utilizing the DeepSeek-R1, we created constant concept anchors and independent concept anchors of different staining agents. During the training phase, we leveraged three types of prompts and a pathological VLM to achieve the description, framing, and reinforcement of the virtual staining direction, thereby optimizing the original virtual staining model. In the inference enhancement phase, we trained learnable denoising prompt blocks based on structural and stylistic constraints, further improving the performance of virtual staining.}
\label{fig2}
\end{figure*}

\section{Related Work}

\subsection{Virtual Staining in Pathology Analysis}

Virtual staining originates from the image-to-image (I2I) task in the domain of natural images, aiming to accomplish the transfer of image styles. With the recent surge in generative model technologies, I2I tasks have also seen significant advancements by leveraging baseline such as GAN~\cite{GAN} and DDPM~\cite{DDPM}. Due to the scarcity and high production costs of paired data, unpaired I2I models have demonstrated greater potential. Zhu \emph{et al.} ~\cite{CycleGAN}designed parallel sets of generators and discriminators to perform staining and restoration on unpaired data separately, achieving commendable results. CUT~\cite{CUT} introduced contrastive learning methods, successfully simplifying the style transfer process by eliminating the need for paired generator-discriminator combinations. Building upon CUT, Jung \emph{et al.} ~\cite{PatchGCL}utilized graph neural networks to identify closely related patches, optimizing the contrastive learning algorithm. Zhao \emph{et al.} ~\cite{egsde}incorporated stochastic differential equations into I2I tasks, achieving a breakthrough in the application of diffusion architectures for unpaired image style transfer. Kim \emph{et al.} ~\cite{UNSB}attempted to leverage Schrödinger bridges to compute the optimal path for domain transfer.

In the field of pathology, virtual staining has demonstrated significant clinical value. A series of H\&E-to-IHC staining workflows facilitate the detection of tumor markers, aiding in tumor classification and diagnosis~\cite{BCI,APT,PPIE,PPT,TDK,PSPsrain}. Virtual staining is employed for tasks such as normalization~\cite{CAGAN,multistain,TTSaD} and tailing artifact reduction~\cite{KIN,FS2FFPE,DN}, enhancing the clarity of tissue sections by mitigating tailing effects and reducing batch effects in pathological slides. Additionally, virtual staining exhibits potential as a proxy task to improve the visual perception capabilities of models~\cite{DiffusionMIL,pathoduet}. Focusing on virtual staining tasks for kidney tissue sections, Lin \emph{et al.} proposed the UMDST~\cite{UMDST} model, which leverages multi-task learning based on stain classification and virtual staining. This model can achieve style transfer across different stains and even simulate virtual scenarios involving multiple stains. Guan \emph{et al.}  ~\cite{GramGAN}designed a style-guided module, showcasing exceptional performance and validating the medical significance of virtual staining in glomerular segmentation and detection tasks. Xiong \emph{et al.} ~\cite{DPI}concentrated on inference enhancement strategies, utilizing the DDIM~\cite{DDIM} architecture to achieve a leap in structural consistency and clarity in virtual staining, thereby addressing the challenges of high-precision medical diagnosis. However, none of these methods incorporate the pathological VLM as a robust assistant to provide enhanced visual recognition capabilities and pathological knowledge, thereby improving the effectiveness of staining.

\subsection{Prompt Tuning in Vision-Language Models}
As of now, the CLIP~\cite{CLIP} architecture remains the mainstream in the VLM and continues to play an irreplaceable role across various visual domains. prompt tuning based on CLIP has also been demonstrated to be an exceedingly simple yet effective strategy. CoOp~\cite{CoOp} and CoCoOp~\cite{CoCoOp} have proven that learnable prompts possess even greater fitting capabilities. Additionally, prompt tuning has achieved exceptional results in tasks such as lighting adjustment~\cite{CLIPlit}, rain and haze removal~\cite{UPIDEDM}, image restoration~\cite{DAclip} and artistic style transfer~\cite{artbank}.

The advancement of prompt tuning methods in the field of pathology is closely intertwined with the maturation of the VLM specialized in pathology. Works such as CONCH~\cite{CONCH}, MUSK~\cite{MUSK}, and PLIP~\cite{plip} have constructed diverse million-scale pathology text-image pairs, demonstrating exceptional performance. Qu \emph{et al.}~\cite{TOP} pioneered the application of prompt tuning on pathological images, enhancing the few-shot classification performance of pathological slides and catalyzing the emergence of outstanding works in subtype classification tasks using prompt Tuning~\cite{Five,VilaMIL,CodMIL,mscpt,fast,focus,QPMIL}. Liu \emph{et al.} ~\cite{vlsa}achieved the first application of prompt tuning in survival analysis by transforming continuous survival labels into textual prompts for ordinal survival learning. To the best of our knowledge, our proposed VPGAN and HARBOR represent the first application of prompt tuning in virtual staining and, more broadly, in medical I2I tasks. Our remarkable performance has been highly encouraging, suggesting that prompt tuning based on the pathology-specific VLM has the potential to become a "super assistant" across all tasks in the field of pathology.

\section{Method}
\subsection{Prompts Generation}
Our method generates different prompts to leverage the pathological knowledge of the VLM for more precise guidance on the staining domain. Moreover, it uses learnable prompts to capture the key information during the virtual staining intermediate process.

\noindent\textbf{Learnable contrastive prompts.}
Contrastive learning methods~\cite{SimCLR,MOCO} often excel in learning the fine-grained features of data such as images. Meanwhile, prompt tuning methods~\cite{CoOp,CoCoOp,CLIPlit,TOP} enhance the accuracy of image text descriptions through a learnable process. Inspired by both, It is proven by us that it is feasible to use the contrastive learning method to capture important information from the intermediate steps of virtual staining and convert it into learnable text prompts during the virtual staining process.

As shown in Figure \ref{fig1}(a) and Figure \ref{fig2}(a) for the training process of contrastive text prompts, we use the training set data from both the source domain images $I_{s}\in {\mathbb{R} }^{H\times W\times 3} $ and the target domain images $I_{t}\in {\mathbb{R} }^{H\times W\times 3} $ of the subsequent staining task as the overall training set. We randomly initialize a positive prompt $T_{p}\in {\mathbb{R} }^{N\times 512}$ and a negative prompt $T_{n}\in {\mathbb{R} }^{N\times 512}$. N
represents the number of embedded tokens in each prompt. Then, we feed the source and target images to the image encoder $\phi _{image}$ of the VLM to obtain their latent code. Meanwhile, we also extract the latent code of the positive and negative prompts by feeding them to the text encoder $\phi _{text}$. Based on the text-image similarity in the VLM latent space, we use the binary cross entropy loss of classifying the source and target images to learn the contrastive prompt pair:
\begin{equation}
\small
\mathcal{L}_{prompt} = -(a*log(\hat{a} )+((1-a)*log(1-\hat{a} ))),\tag{1}
\end{equation}
\begin{equation}
\small
\hat{a} = \frac{e^{\cos(\Phi_{image}(I), \Phi_{text}(T_p))}}{\sum_{i \in \{n, p\}} e^{\cos(\Phi_{image}(I), \Phi_{text}(T_i))}},\tag{2}
\end{equation}
where $I\in \left \{ I_s,I_t \right \} $ and $a$ is the label of the current image, 0
is for negative sample $I_s$ and 1 is for positive sample $I_t$.

\noindent\textbf{Concept Anchors.} 
In this section, we aim to generate pathological visual descriptions to serve as prior linguistic knowledge for guiding the virtual staining task. To minimize manual effort, large language models (LLMs)~\cite{GPT4,deepseekR1} are employed to produce descriptions related to different staining agents. Specifically, we input the following query into the LLM: \textit{"We want to study the effects of different staining agents on the same liver pathological section. We will use the histological images of human kidney at 40x magnification. Please tell me the visual characteristics of {Class} staining compared with other staining agents and the key observation areas."} Following a similar approach, we obtain the intrinsic feature description sets for kidney tissue sections, thereby deriving a total of five concept knowledge sets corresponding to the four staining classes and the intrinsic features. Ultimately, we utilized the text encoder $\phi _{text}$ of VLM to generate the final concept anchors.

\subsection{VLM-based Prompts GAN}
Building upon the aforementioned learnable contrastive prompts, the invariant concept anchors of kidney tissue sections, and the independent concept anchors of different staining agents, we propose the \textbf{V}LM-based \textbf{P}rompts \textbf{G}enerative \textbf{A}dversarial \textbf{N}etwork (VPGAN). This framework enhances the original GAN~\cite{GAN} architecture for unpaired data by meticulously characterizing the intermediate staining processes, the fixed concepts of kidney tissue, and the staining agent-specific concepts, thereby improving the overall image generation quality. Furthermore, our VPGAN is adaptable to any GAN architecture as an optimization method. After extensive experimentation, we selected CycleGAN~\cite{CycleGAN} as the baseline for this paper, thereby achieving the optimal generation results. Apart from our design, the configurations for the generator, discriminator, and learning rate are all aligned with the original settings of CycleGAN.

\noindent\textbf{Contrastive Prompt Transfer Module.} 
Inspired by CLIP-LIT~\cite{CLIPlit}, we prove that learnable contrastive prompts can achieve effective image enhancement. Building on this, we introduce the Contrastive Prompt Transfer Module (CPT), which decodes pathological information from the VLM to describe the intermediate staining processes and subtle domain-specific differences in staining. To the best of our knowledge, this represents the first application of learnable contrastive prompts in GAN models.

Given the learnable contrastive prompts obtained from the prompts generation step, we can train the CPT module with VLM-aware loss. This loss is based on the contrastive differences between staining domains and depicts the staining transfer process, thereby improving the quality of virtual staining.
\begin{equation}
\small
\mathcal{L}  _{contrastive} = \frac{e^{\cos(\Phi_{image}(I), \Phi_{text}(T_n))}}{\sum_{i \in \{n, p\}} e^{\cos(\Phi_{image}(I),\Phi_{text}(T_i))}}.\tag{3}
\end{equation}

\noindent\textbf{Constant Concept Anchoring Module.} 
In fact, the most significant difference between virtual staining and natural image style transfer lies in the need to consider the preservation of texture and shape, the authenticity of pathological features, and the real physicochemical properties of staining agents. Merely achieving the highest degree of fitting in terms of style will lead to a substantial reduction in clinical efficacy. This is also the crucial reason why a large number of evaluation metrics for virtual staining tasks place greater emphasis on structural preservation~\cite{UMDST,DPI,KIN}. In addition, the inherent properties in imaging, such as the magnification of the slices and the imaging equipment used for the dataset, are also taken into account. As described in the prompt generation, we leverage the powerful capabilities of DeepSeek-R1~\cite{deepseekR1} and online searches to obtain relatively accurate descriptions, and manually remove parts with factual errors caused by LLM hallucinations.

Next, the goal of Constant Concept Anchoring Module (CCA) is to quantify the invariance between the pre-staining images $I_{pre}\in {\mathbb{R} }^{H\times W\times 3} $ and the post-staining images $I_{post}\in {\mathbb{R} }^{H\times W\times 3} $. To this end, we generate the concept of renal slice invariance, denoted as ${\mathbb{R} }^{1\times 512}$, which is derived by first generating conceptual descriptions via the LLM and subsequently transforming them into textual embeddings through the VLM. We use CPT's method to measure image-text correspondence with cosine similarity, obtaining cosine similarities $S_{pre}$ and $S_{post}$ for subsequent concept invariance analysis. The formula is as follows:
\begin{equation}
\small
S_{pre}=e^{\cos(\Phi_{image}(I_{pre}), \Phi_{text}(T_{constant}))},\tag{4}
\end{equation}
\begin{equation}
\small
S_{post}=e^{\cos(\Phi_{image}(I_{post}), \Phi_{text}(T_{constant}))}.\tag{5}
\end{equation}
Subsequently, to ensure the invariance of the constant concept before and after staining, we employed the Mean Squared Error (MSE) loss function to calculate the mean of the sum of squared differences between the cosine similarities $S_{pre}$ and $S_{post}$ before and after staining, thereby completing the delineation of the virtual staining range. The formula for the constant concept loss function $\mathcal{L} _{constant}$ is as follows:
\begin{equation}
\small
\mathcal{L} _{constant}=\left | \left | S_{pre}-S_{post} \right |  \right |_{2}  .\tag{6}
\end{equation}

\noindent\textbf{Independent Concept Reinforcement Module.} 
The staining effects are aimed to be further enhanced by us by leveraging the independent concept anchors of staining agents generated by LLM. Based on textual prompts from VLMs, more microscopic details that are difficult to capture in conventional GANs can be obtained, such as the staining effects of specific agents on cell nuclei in pathological slides. It is important to emphasize that certain "shortcut" prompts that quickly deceive the discriminator (\emph{e.g.} black streaks produced by PASM staining) will be eliminated, as their use would undermine our goal of achieving fine-grained textual prompts and instead exacerbate overfitting in style transfer. The final selected textual prompts for the four staining agents are collectively represented as ${\mathbb{R} }^{4\times 512}$ after being encoded into text embeddings by the VLM. Subsequently, we compute the cosine similarity $S_{stain}$ between the stained image $I_{post}$ and the textual prompts $T_{stain}$ containing information about the four types of staining agents (H\&E, MAS, PAS, PASM). Based on the inherent preprocessing method and computational rules of the CE loss, we perform a four-class proxy task to ensure that the stained image $I_{post}$ approximates the target staining domain as closely as possible. The formula is as follows:
\begin{equation}
\small
S_{stain}=e^{\cos(\Phi_{image}(I_{post}), \Phi_{text}(T_{stain}))},\tag{7}
\end{equation}
\begin{equation}
\small
p= softmax(S_{stain}),\tag{8}
\end{equation}
\begin{equation}
\small
\mathcal{L} _{stain}= -\frac{1}{m} \sum_{i=0}^{m} (one-hot)_{p_i} \cdot log(p_i),\tag{9}
\end{equation}
where $p$ represents the result of normalizing the similarity $S_{stain}$. We use $one-hot$ encoding to describe the category of the virtually stained image, which is consistent with the category of the target staining domain. In summary, the loss function of VPGAN is as follows, where $\alpha$, $\beta$, and $\gamma$ are the hyperparameters therein:
\begin{equation}
\small
\mathcal{L}_{train}= \mathcal{L}_{normal}+\alpha   \mathcal{L}_{contrastive}+\beta \mathcal{L}_{constant} + \gamma\mathcal{L}_{stain} .
\tag{10}
\end{equation}

\begin{figure}[tp]
 \includegraphics[width=0.95\linewidth]{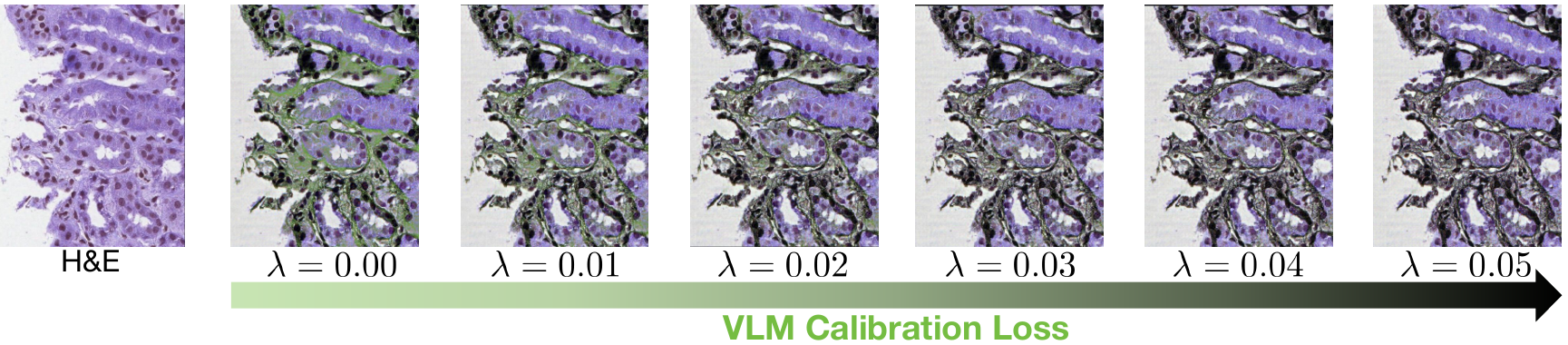}
\caption{We demonstrate a fine-grained verification process based on the VLM on the H\&E2PASM task, enabling the progressive and successful verification of the staining domains.}
\label{fig3}
\end{figure}

\subsection{Inference Enhancement}
Dual-Path Inference (DPI) \cite{DPI}, based on DDIM~\cite{DDIM}, achieves inference enhancement. It has been demonstrated that this method improves the integrity of pathological tissue structures and significantly reduces the distortion of virtually stained images. However, it carries a substantial risk of staining domain collapse. To address this issue, we introduce fine-grained structural verification based on the VLM, which successfully resolves the problem of staining domain collapse and further enhances performance. The progressive correction effects are illustrated in Figure \ref{fig3}.

\noindent\textbf{Inference Enhancement Baseline.} 
We adopted the same settings as DPI and pre-trained a DDIM with a step size of 50. As a result, we can obtain the noisy image $\mathcal{X}_{t_k}$ at the $k$-th step based on DDIM for the original image $I_{pre}$. Similarly, we can get the intermediate noisy image $\mathcal{Y}_{t_k}$ for the image $I_{post}$ after virtual staining by VPGAN. The recurrence formulas for $\mathcal{X} = \left \{\mathcal{X}_{t_1}, \mathcal{X}_{t_2}, \cdot \cdot \cdot , \mathcal{X}_{t_k},\cdot \cdot \cdot,\mathcal{X}_{t_{50}}  \right \} $ and $\mathcal{Y} = \left \{\mathcal{Y}_{t_1}, \mathcal{Y}_{t_2}, \cdot \cdot \cdot , \mathcal{Y}_{t_k},\cdot \cdot \cdot,\mathcal{Y}_{t_{50}}  \right \} $ in DDIM are as follows:
\begin{equation}
\small
\mathcal{X}_{t_{k+1}} = \sqrt{\alpha_{k+1}} \left( \frac{\mathcal{X}_{t_{k}} - \sqrt{1 - \alpha_k} \epsilon_\theta(\mathcal{X}_{t_{k}}, k, C_S)}{\sqrt{\alpha_k}} \right) + \sqrt{1 - \alpha_{k+1}} \epsilon_\theta(\mathcal{X}_{t_{k}}, k, C_S),\tag{11}
\end{equation}
\begin{equation}
\small
\mathcal{Y}_{t_{k+1}} = \sqrt{\alpha_{k+1}} \left( \frac{\mathcal{Y}_{t_{k}} - \sqrt{1 - \alpha_k} \epsilon_\theta(\mathcal{Y}_{t_{k}}, k, C_E)}{\sqrt{\alpha_k}} \right) + \sqrt{1 - \alpha_{k+1}} \epsilon_\theta(\mathcal{Y}_{t_{k}}, k, C_E).\tag{12}
\end{equation}
The diffusion model’s conditional variables are denoted as $C_S$ and $C_T$, where $C_S$ represents the source domain category conditional variable, and $C_E$ represents signifies the absence of additional conditional variables to mitigate errors in the style trajectory. $\epsilon_\theta$ is a neural network controlled by the parameter $\theta$, and the main function of this network is to predict noise.

Our objective is to augment each stained image by in proper order training $\mathcal{Z} = \left \{\mathcal{Z}_{t_1}, \mathcal{Z}_{t_2}, \cdot \cdot \cdot , \mathcal{Z}_{t_k},\cdot \cdot \cdot,\mathcal{Z}_{t_{50}}  \right \} $, an initially zero-initialized empty prompt map, which serves as an additional noise prompt to achieve the effect of data augmentation. $\mathcal{Z}$ is trained based on the SSIM structural constraint of $\mathcal{X}$ and the MSE stylization constraint of $\mathcal{Y}$. The formula is as follows:
\begin{equation}
\small
\mathcal{L}_{struct} = \sum_{k=0}^{50} \frac{2(\sigma_{\mathcal{Z}_{t_k}}\sigma_{\mathcal{X}_{t_k}} + c_1) \times (\sigma_{\mathcal{Z}_{t_k} \mathcal{X}_{t_k}} + c_2)}{(\sigma_{\mathcal{Z}_{t_k}}^2 + \sigma_{\mathcal{X}_{t_k}}^2 + c_1)(\sigma_{\mathcal{Z}_{t_k}}\sigma_{\mathcal{X}_{t_k}} + c_2)}
,\tag{13}
\end{equation}
\begin{equation}
\small
\mathcal{L}_{style } = \sum_{k=0}^{50} \left | \left | \mathcal{Z}_{t_k}-\mathcal{Y}_{t_k} \right |  \right |_{2} .\tag{14}
\end{equation}
Among them, $\sigma_{\mathcal{X}_{t_k}}$ and $\sigma_{\mathcal{Z}_{t_k}}$ are the variances of $\mathcal{X}_{t_k}$ and $\mathcal{Z}_{t_k}$, $\sigma_{\mathcal{Z}_{t_k}\mathcal{X}_{t_k}}$ is their covariance.

We denoise and restore $\mathcal{Y}$ based on the cues from $\mathcal{Z}$ to obtain the enhanced image $I_{enhance}\in {\mathbb{R} }^{H\times W\times 3} $. The formula is as follows:
\begin{equation}
\small
\mathcal{Y}_{t_k}^*= \mathcal{Y}_{t_k}+\mathcal{Z}_{t_k}
,\tag{15}
\end{equation}
\begin{equation}
\begin{aligned}
\small
\psi (\mathcal{Y}_{t_{k-1}}^*,C_T) &= \sqrt{\alpha_{k - 1}} \left( \frac{\mathcal{Y}_{t_{k}}^* - \sqrt{1 - \alpha_k} \epsilon_\theta(\mathcal{Y}_{t_{k}}^*, k, C_T)}{\sqrt{\alpha_k}} \right) \\
&+ \sqrt{1 - \alpha_{k - 1}} \epsilon_\theta(\mathcal{Y}_{t_{k}}^*, k, C_T),
\end{aligned}
\tag{16}
\end{equation}
where the $\psi$ function represents conditional sampling, and $C_T$ represents the target domain label. Finally, $I_{enhance}$ is obtained.

\begin{table*}[htbp]
    \centering
    \caption{Comparison of different methods on H\&E2MAS,H\&E2PAS and H\&E2PASM datasets. The parts with a \textcolor{red!30}{red background} represent zero-cost inference methods, and the best results are marked in \textcolor{red}{red}. The parts with a \textcolor{blue!30}{blue background} represent inference enhancement methods, and the best results are marked in \textcolor{blue}{blue}.}
    \resizebox{0.95\textwidth}{!}{
    \begin{tabular}{l | c c c c c | c c c c c | c c c c c}
        \toprule
        \multirow{2}{*}{Method} & \multicolumn{5}{c}{H\&E2MAS} & \multicolumn{5}{c}{H\&E2PAS}&\multicolumn{5}{c}{H\&E2PASM} \\
        \cmidrule(lr){2-6} \cmidrule(lr){7-11}\cmidrule(lr){12-16}
        & SSIM$\uparrow$ & CSS$\uparrow$ & MS-SSIM$\uparrow$ & PSNR$\uparrow$ & FID & SSIM$\uparrow$ & CSS$\uparrow$ & MS-SSIM$\uparrow$ & PSNR$\uparrow$ & FID& SSIM$\uparrow$ & CSS$\uparrow$ & MS-SSIM$\uparrow$ & PSNR$\uparrow$ & FID \\
        \midrule
        \rowcolor{red!5}CycleGAN~\cite{CycleGAN} &0.7824&0.8336&0.7843&15.71&133.17&0.8654&0.8888&0.8914&17.12&149.81&0.5490&0.6391&0.4013&10.54&140.44\\
        \rowcolor{red!5}CUT~\cite{CUT}&0.7075&0.7519&0.8090&16.03&125.64&0.7157&0.7337&0.8496&16.65&103.02&0.2914&0.3180&0.3293&10.14&103.96  \\
        
        \rowcolor{red!5}UGATIT~\cite{UGATIT}&0.6091&0.6322&0.8412&16.24&152.48&0.5658&0.5793&0.8542&16.70&111.04&0.3680& 0.4013&0.3024&10.96&136.72 \\
        \rowcolor{red!5}UNSB~\cite{UNSB}&0.6224&0.6773&0.7823&14.38&120.13&0.6648&0.6752&0.8701&18.3&112.46&0.2849&0.3119&0.2938&10.39&\textcolor{red}{87.54}  \\
        \rowcolor{red!5}PatchGCL~\cite{PatchGCL}&0.3677&0.4199&0.6777&11.99&123.73&0.4940&0.5039&0.8011&16.46&\textcolor{red}{91.99}& 0.2346&0.2573&0.2623&10.04&95.37 \\
        \rowcolor{red!5}GramGAN~\cite{GramGAN}&0.6260&0.6767&0.8073&14.77&175.83&0.6938&0.7096&0.8739&17.12&154.06&0.5088&0.5653&0.5327&12.17& 174.81 \\
        \rowcolor{red!5}UMDST~\cite{UMDST} &0.7514&0.7864&0.8571&\textcolor{red}{17.16}&187.99&0.7762&0.7958&0.9254&17.12&154.06&0.5845&0.6276&0.5432&12.23&129.59\\
        \rowcolor{red!5}VPGAN(Ours)&\textcolor{red}{0.8158}&\textcolor{red}{0.8648}&\textcolor{red}{0.8526}&16.49&\textcolor{red}{112.06}&\textcolor{red}{0.9173}&\textcolor{red}{0.9339}&\textcolor{red}{0.9457}&\textcolor{red}{19.06}&132.95&\textcolor{red}{0.6650}&\textcolor{red}{0.7372}&\textcolor{red}{0.5997}&\textcolor{red}{12.65}&125.28 \\
        \midrule
        \rowcolor{blue!5} DPI~\cite{DPI}&0.8971&0.9040&0.9278&20.86&193.94&0.8935&0.8883&0.9508&22.25&157.47&——&——&——&——&——  \\
        \rowcolor{blue!5} HARBOR(Ours)& \textcolor{blue}{0.9063}& \textcolor{blue}{0.9149}& \textcolor{blue}{0.9312}& \textcolor{blue}{21.02}& \textcolor{blue}{152.94}& \textcolor{blue}{0.9302}& \textcolor{blue}{0.9343}& \textcolor{blue}{0.9643}& \textcolor{blue}{23.64}& \textcolor{blue}{154.09}& \textcolor{blue}{0.6736}& \textcolor{blue}{0.7323}& \textcolor{blue}{0.6498}& \textcolor{blue}{13.30}& \textcolor{blue}{132.77} \\
        \bottomrule
    \end{tabular}
}
\label{tab1}
\end{table*}

\begin{figure*}[tp]
 \includegraphics[width=\linewidth]{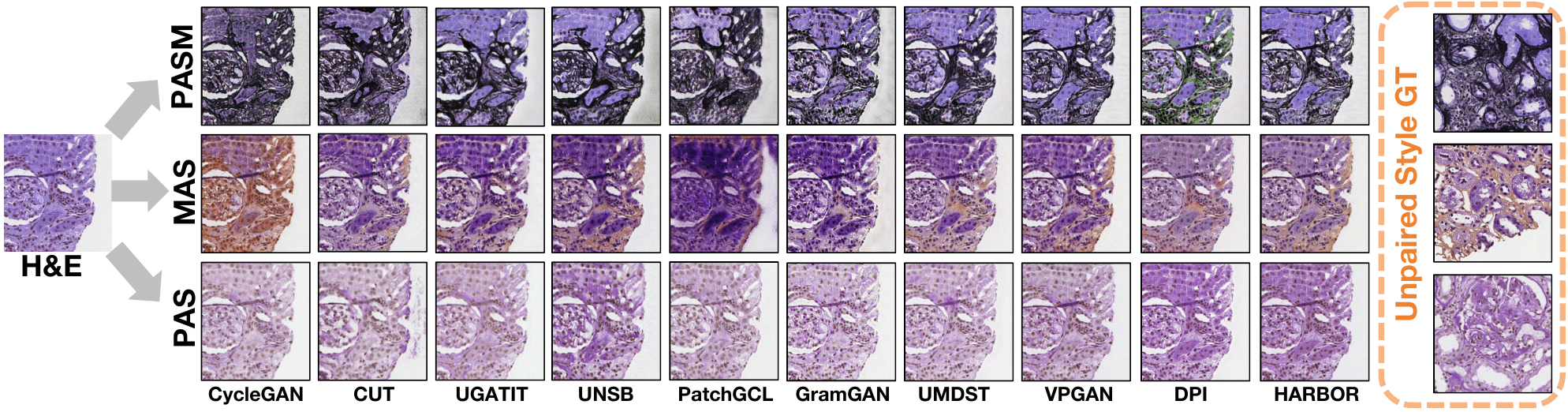}
\caption{The performance comparison of various existing methods and our proposed method for multiple stain transfer of the same H\&E-stained image.}
\label{fig4}
\end{figure*}

\noindent\textbf{Structural Verification based on the VLM.} 
Without altering the DDIM-based inference enhancement framework, we aim to leverage the powerful capabilities of VLMs to address the issue of staining domain degradation. Since the focus here is primarily on correcting structural and color-related aspects, we employ CLIP~\cite{CLIP}, a general-purpose VLM, rather than pathology-specialized VLMs such as CONCH. For the visual encoder $\varphi _{image}$, we utilize ResNet101~\cite{resnet}, which allows us to extract intermediate layer features for detailed corrections, thereby resolving staining domain degradation and further enhancing the robustness and performance of the inference enhancement. The loss function is as follows:
\begin{equation}
\small
\mathcal{L}_{calibration} = \sum_{k=0}^{50}\sum_{l=0}^{4} \delta_l \cdot \left\| \varphi _{image}^l(\mathcal{Z}_{t_k}) - \varphi _{image}^l(\mathcal{Y}_{t_k}) \right\|_2,
\tag{17}
\end{equation}
where $\delta_l$ is the weight of the $l$-th layer of the image encoder in the ResNet101 CLIP model. Finally, our inference-enhanced loss function is as follows:
\begin{equation}
\small
\mathcal{L}_{enhance}= \mu  \mathcal{L}_{struct}+(1- \mu)\mathcal{L}_{style} + \lambda\mathcal{L}_{calibration} ,
\tag{18}
\end{equation}
$\mu$ and $\lambda$ all adjustable parameters. Details of hyperparameter settings in our method can be found in the \textbf{supplementary materials}.

\section{Experiments and Results}

\begin{figure}[tp]
 \includegraphics[width=0.6\linewidth]{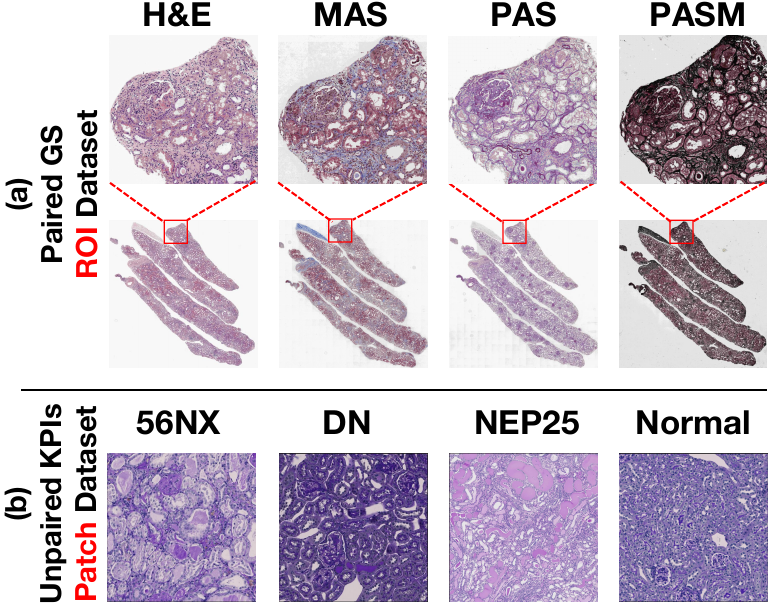}
\caption{Overview of Downstream Task Datasets}
\label{fig6}
\end{figure}

\subsection{Datasets and Experiment Setup}
\noindent\textbf{Datasets.} 
As depicted in Figure \ref{fig6}, we evaluated the performance of our method on three open-source datasets. Following the settings of UMDST~\cite{UMDST} and DPI~\cite{DPI}, we obtained and partitioned ANHIR dataset~\cite{anhir} slices stained with H\&E, MAS, PAS, and PASM, enabling model training and performance validation. On the GS~\cite{GramGAN} dataset, the virtual staining effect on glomerular detection and segmentation at the ROI level was examined by us adhering to the GramGAN setup. For the KPIs dataset~\cite{kpis}, we normalized PAS slices across different disease categories according to its data settings, thereby enhancing the performance of glomerular object detection at the patch level. Please refer to the \textbf{supplementary materials} for all specific details of dataset division and preprocessing.

\noindent\textbf{Evaluation Metrics.} 
In this experiment, we employed five metrics to comprehensively evaluate the performance of pathological image translation. First, the Structural Similarity Index \textbf{(SSIM)} was used to measure the similarity in luminance, contrast, and structure between images. Second, the Contrast Structural Similarity \textbf{(CSS)} focused on assessing the preservation of contrast and structural details. Additionally, the Multi-Scale Structural Similarity \textbf{(MS-SSIM)} evaluated both fine details and global structures through multi-scale analysis. Meanwhile, the Peak Signal-to-Noise Ratio \textbf{(PSNR)} quantified the noise and distortion levels in the images. Finally, the Fréchet Inception Distance \textbf{(FID)} assessed the distribution consistency between generated and real images in the feature space. These metrics ensured a thorough and reliable evaluation of the results from multiple perspectives. Our approach achieved state-of-the-art (SOTA) performance across multiple metrics, validating the effectiveness of the proposed method.

\begin{table*}[htbp]
    \centering
    \caption{An ablation study was conducted on the CPT, CCA, and ICR modules in the VPGAN. All the research was carried out on H\&E2MAS, H\&E2PAS, and H\&E2PASM. The best values are highlighted.}
    \resizebox{0.95\textwidth}{!}{
    \begin{tabular}{c c c | c c c c c | c c c c c | c c c c c} 
        \toprule
        \multicolumn{3}{c|}{Module} & \multicolumn{5}{c}{H\&E2MAS} & \multicolumn{5}{c}{H\&E2PAS} & \multicolumn{5}{c}{H\&E2PASM} \\
        \cmidrule(lr){1-3} \cmidrule(lr){4-8} \cmidrule(lr){9-13} \cmidrule(lr){14-18}
        CPT & CCA & ICR & SSIM$\uparrow$ & CSS$\uparrow$ & MS-SSIM$\uparrow$ & PSNR$\uparrow$ & FID & SSIM$\uparrow$ & CSS$\uparrow$ & MS-SSIM$\uparrow$ & PSNR$\uparrow$ & FID & SSIM$\uparrow$ & CSS$\uparrow$ & MS-SSIM$\uparrow$ & PSNR$\uparrow$ & FID \\
        \midrule
        - & - & - &0.7824&0.8336&0.7843&15.71&133.17&0.8654&0.8888&0.8914&17.12&149.81&0.5490&0.6391&0.4013&10.54&140.44\\

        $\checkmark$ & - & - &0.7964&0.8574&0.8335&15.35&121.83&0.8729&0.8926&0.8907&18.52&143.20&0.5447&0.6422&0.4342&10.27&138.88\\

        - & $\checkmark$ & - &0.7686&0.8375&0.8321&14.80&124.10&0.8601&0.8813&0.8931&17.92&147.93&0.5236&0.6144&0.3732&10.50&140.06\\

        - & - & $\checkmark$ &0.7599&0.8310&0.8180&14.66&\textbf{109.41}&0.9031&0.9234&0.9219&18.27&134.02&0.5858&0.6541&0.4214&11.51&143.79\\

        $\checkmark$ & $\checkmark$ & - &0.7763&0.8618&0.8411&15.80&135.45&0.8839&0.8812&0.8923&18.84&153.31&0.6278&0.6972&\textbf{0.6069}&11.97&151.22\\

        $\checkmark$ & - & $\checkmark$ &0.7993&0.8520&0.8416&15.59&122.43&0.9102&0.9081&0.9414&18.02&140.16&0.6495&0.7031&0.5734&12.57&127.65\\

        - & $\checkmark$ & $\checkmark$ &0.7450&0.8061&0.7983&15.28&113.73&0.8592&0.8932&0.8864&17.31&137.84&0.5901&0.6528&0.4370&11.52&132.18\\

        $\checkmark$ & $\checkmark$ & $\checkmark$& \textbf{0.8158}&\textbf{0.8648}&\textbf{0.8526}&\textbf{16.49}&112.06&\textbf{0.9173}&\textbf{0.9339}&\textbf{0.9457}&\textbf{19.06}&\textbf{132.95}&\textbf{0.6650}&\textbf{0.7372}&0.5997&\textbf{12.65}&\textbf{125.28} \\
        \bottomrule
    \end{tabular}
}
\label{tab2}
\end{table*}

\begin{table*}[htbp]
    \centering
    \caption{Comparison of different VLMs' effect on H\&E2MAS, H\&E2PAS and H\&E2PASM datasets. The best values are highlighted.}
    \resizebox{0.95\textwidth}{!}{
    \begin{tabular}{l | c c c c c | c c c c c | c c c c c}
        \toprule
        \multirow{2}{*}{VLM} & \multicolumn{5}{c}{H\&E2MAS} & \multicolumn{5}{c}{H\&E2PAS}&\multicolumn{5}{c}{H\&E2PASM} \\
        \cmidrule(lr){2-6} \cmidrule(lr){7-11}\cmidrule(lr){12-16}
        & SSIM$\uparrow$ & CSS$\uparrow$ & MS-SSIM$\uparrow$ & PSNR$\uparrow$ & FID & SSIM$\uparrow$ & CSS$\uparrow$ & MS-SSIM$\uparrow$ & PSNR$\uparrow$ & FID& SSIM$\uparrow$ & CSS$\uparrow$ & MS-SSIM$\uparrow$ & PSNR$\uparrow$ & FID \\
        \midrule
        CLIP~\cite{CLIP} &0.7215&0.7531&0.7309&16.20&155.34&0.8336&0.8528&0.8817&17.04&145.98&0.5827&0.6465&0.4835&11.89&143.08\\
        PLIP ~\cite{plip}&0.7776&0.8567&0.8217&15.12&114.03&0.8745&0.8866&0.9000&18.23&141.29&0.6591&0.6922&0.5075&10.93&139.67\\
        MUSK ~\cite{MUSK}&\textbf{0.8259}&0.8624&\textbf{0.8670}&\textbf{16.71}&130.43&0.8854&0.8871&0.9202&18.95&153.22&0.6437&0.7052&\textbf{0.6214}&12.42&127.42\\
        CONCH ~\cite{CONCH}
        &0.8158&\textbf{0.8648}&0.8526&16.49&\textbf{112.06}&\textbf{0.9173}&\textbf{0.9339}&\textbf{0.9457}&\textbf{19.06}&\textbf{132.95}&\textbf{0.6650}&\textbf{0.7372}&0.5997&\textbf{12.65}&\textbf{125.28} \\
        \bottomrule
    \end{tabular}
}
\label{tab3}
\end{table*}

\noindent\textbf{Implementation Details.} 
Our method is implemented in PyTorch and trained on a workstation with 8 NVIDIA H100 GPUs. We adopted CycleGAN~\cite{CycleGAN} as the baseline for training VPGAN, setting the batch size to 1 and training for 50 epochs. During the inference enhancement phase, we utilized DPI~\cite{DPI} as the baseline. The data preprocessing methods, optimizer, and learning rate settings were kept consistent with the baseline. The remaining hyperparameter configurations are detailed in the \textbf{supplementary materials}.

\subsection{Comparison Results} VPGAN and HARBOR were compared with previous unpaired image translation and virtual staining methods by us. Given the substantial computational cost and time overhead of the DDIM-based inference enhancement technique (requiring 5-10 minutes for inference enhancement on a single $256\times256$ image), we deemed it necessary to separately evaluate zero-cost inference methods and inference enhancement methods. This approach further validates the versatility of our method across diverse scenarios.

\noindent\textbf{Zero-Cost Inference Methods.} 
We selected \textbf{CycleGAN} as the baseline for \textbf{VPGAN} and compared it with various two-domain unpaired image translation methods such as \textbf{CUT}~\cite{CUT}, \textbf{UGATIT}~\cite{UGATIT}, \textbf{UNSB}~\cite{UNSB}, and \textbf{PatchGCL}~\cite{PatchGCL}, as well as multi-domain unpaired image translation methods like \textbf{GramGAN} and \textbf{UMDST}. As shown in Table \ref{tab1} and Figure \ref{fig4}, VPGAN achieved SOTA performance across 12 metrics on three datasets compared to other zero-cost inference methods, and also demonstrated the best visual quality. It is clearly evident that compared to the baseline CycleGAN, our method effectively addresses its limitations in the transitional style transfer for the H\&E2PASM and H\&E2MAS tasks. By leveraging VLM-based prompt constraints, VPGAN achieves remarkable results that faithfully adhere to the staining characteristics of pathological images. Both CUT and PatchGCL demonstrate structural artifacts, and the PatchGCL method entirely collapses in the H\&E2MAS task. GramGAN exhibits noticeable blurring at the edges of the images and irregular stains in the H\&E2PASM task, which contradicts the physical properties of the staining agents. In contrast, UNSB and UMDST perform slightly worse in terms of image clarity and structural preservation. The diversity of issues encountered with other methods underscores the versatility of VLMs as a virtual staining assistant, playing a significant role in various aspects such as pathological knowledge guidance and structural preservation.

\noindent\textbf{Inference Enhancement Methods.} 
Our method, HARBOR, in comparison to the baseline DPI, has further rectified the deviation in the style domain. In the H\&E2PASM task, it successfully repaired the complete collapse of the DPI staining domain (manifested by the emergence of irrelevant green colors), and in the H\&E2MAS task, it also addressed the issue of green residual shadows in the cell nuclei. In addition, visual calibration based on the VLM also enables HARBOR to achieve the SOTA performance in all indicators among inference enhancement methods. Compared to VPGAN, HARBOR better demonstrates the texture and veins of the images. This is due to the original images' texture cues and VLMs' strong visual discrimination. This may be the underlying reason for its superior performance over VPGAN across multiple metrics.

\subsection{Ablation Study on VPGAN} As shown in Table \ref{tab2} and Table \ref{tab3}, we conducted a series of ablation studies on each module of VPGAN and the underlying VLM assistants. These studies demonstrated the necessity of each module and enabled the selection of the Best VLM for the staining task. 

\noindent\textbf{Module Ablation of VPGAN.} 
We investigated the functions and necessity of the CPT, CCA, and ICR modules in VPGAN, as shown in Table \ref{tab2}. Excitingly, the results on three datasets indicate that the functions of these modules are complementary rather than simply a superposition of performance. It can be observed that the enhancement of fixed style domains is achieved by the ICR module, which is also reflected in the general improvement of the FID metric. The CPT module is capable of representing complex intermediate coloring processes, leading to performance enhancements across multiple metrics. The CCA module, when used alone for structural invariance correction, may even yield results inferior to the baseline. However, when combined with other modules that describe the coloring process and reinforce specific coloring domains, it collectively achieves superior outcomes. This conclusion is quite intriguing. Drawing an analogy to our daily lives, an assistant who only points out what cannot be done might be quite frustrating. Yet, after summarizing the specific workflow and key tasks, appropriate regulations and reminders can further enhance work efficiency.

\noindent\textbf{Performance Differences across VLMs.} 
Due to differences in data sources, data volume, and training methods, VLMs may exhibit performance variations in virtual staining tasks. We conducted a comparative analysis of the effectiveness of CLIP~\cite{CLIP} on natural images and pathology-specialized models such as PLIP~\cite{plip}, MUSK~\cite{MUSK}, and CONCH~\cite{CONCH} on VPGAN across three datasets: H\&E2MAS, H\&E2PAS, and H\&E2PASM. As shown in Table \ref{tab3}, the experimental results demonstrate that CONCH achieves the optimal performance on VPGAN. Based on the experimental results, we observed that CLIP even leads to a performance degradation compared to the baseline, which may stem from its inherent incompatibility with medical texts and staining tasks. The performance differences between PLIP and other pathology-specialized VLMs are attributed to its slightly inferior data volume and quality, which is also reflected in VLM-based subtype classification and survival analysis tasks. CONCH and MUSK demonstrated comparable results, but CONCH exhibited more balanced outcomes, likely due to its training data sources. Consequently, we selected CONCH as the VLM for our method, as it achieved the best average performance.

\begin{table}[tp]
\centering
\caption{Search for the optimal value of the hyperparameter $\lambda$ for calibration in the H\&E2MAS task. }
\resizebox{0.45\textwidth}{!}{%
\begin{tabular}{l|ccccc}
\toprule
$\lambda$ setting & SSIM$\uparrow$ & CSS$\uparrow$ & MS - SSIM$\uparrow$ & PSNR$\uparrow$ & FID \\
\midrule
$\lambda=0.00$ & 0.8794 & 0.8942 & 0.9254 & 19.86 & 197.52 \\
$\lambda=0.0005$ & 0.8797 & 0.9023 & 0.9231 & 20.12 & 173.20 \\
$\lambda=0.001$ & \textbf{0.9063} & \textbf{0.9149} & \textbf{0.9312} & \textbf{21.02} & 152.94 \\
$\lambda=0.005$ & 0.8925 & 0.9010 & 0.9219 & 20.83 & 147.94 \\
$\lambda=0.01$ & 0.8738 & 0.9048 & 0.9283 & 20.07 & \textbf{142.83} \\
$\lambda=0.05$ & 0.8682 & 0.8700 & 0.8928 & 19.53 & 230.78 \\
\bottomrule
\end{tabular}
}
\label{tab4}
\end{table}

\begin{figure}[tp]
 \includegraphics[width=0.95\linewidth]{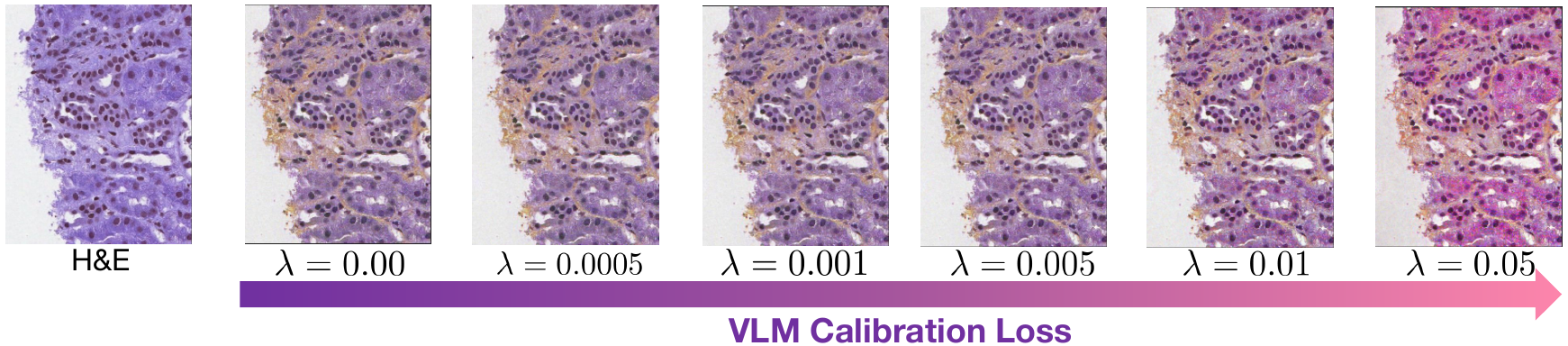}
\caption{Unlike H\&E2PASM, H\&E2MAS correction shows post-staining domain re-collapse from over-correction, necessitating exploration of the optimal correction interval.}
\label{fig5}
\end{figure}

\subsection{The Optimal Interval of the Calibration}

In contrast to the reasoning enhancement component in H\&E2PASM, we observed that on H\&E2MAS, reasoning enhancement exhibits a more pronounced effect within a certain range of the correction loss function compared to H\&E2PAS and H\&E2PASM tasks. However, when the constraint properties become excessively strong, it leads to a secondary collapse in the staining domain, resulting in an unrealistic bright pink color. This intriguing phenomenon necessitates the search for the optimal hyperparameter $\lambda$ in the correction loss function on the H\&E2MAS dataset, guiding the reasoning enhancement to the optimal staining range.

Table \ref{tab4} demonstrates the effects of reasoning enhancement under different $\lambda$ settings in the H\&E2MAS task. It can be observed that, within a certain parameter range, the calibration loss function effectively corrects the collapse in the style domain, which is also reflected in the continuous decrease of the FID score. At $\lambda=0.001$, the optimal structural correction result is achieved, along with a style correction result that approximates the optimal outcome. The optimal range for correction is derived from our experiments. Interestingly, when the $\lambda$ used for correcting the color domain is greater than or equal to $0.05$, another type of collapse in the color domain occurs. This indicates that the correction process based on the VLM is dynamic, and granting the "assistant" too much power can lead to disastrous results.

\subsection{Downstream Tasks}
Our model's superior performance in multi-scale glomerular detection and segmentation on ROI and patch-level datasets was validated. Due to significant color and pathological differences among DS, KPIs, and ANINR datasets, we trained and tested each separately, maintaining consistent preprocessing and training. Given DDIM's size constraints, we conducted downstream experiments only on VPGAN, showcasing our method's clinical value.

\begin{table}[tp]
    \centering
    \caption{We use MAP@[0.50:0.95] to evaluate the accuracy of ROI-level glomeruli detection and segmentation.}
    \resizebox{0.45\textwidth}{!}{
    \begin{tabular}{l | c c c c}
        \toprule
        \multirow{2}{*}{Tasks} & H\&E & PASM & PAS & MAS \\
        & (real) & (generated) & (generated) & (generated) \\
        \midrule
        Detection & 0.543 & \textbf{0.559} & 0.548 & 0.546 \\
        Segmentation & 0.567 & \textbf{0.581} & 0.574 & 0.528 \\
        \bottomrule
    \end{tabular}
    }
    \label{tab5}
\end{table}

\begin{table}[tp]
\centering
\caption{Patch-level glomerulus segmentation accuracy.}
\resizebox{0.45\textwidth}{!}{%
\begin{tabular}{l|ccc}
\toprule
\textbf{Method} &  \textbf{Average} & \textbf{Merge} & \textbf{Normalization by VPGAN}\\
\midrule
Unet~\cite{unet}  & 87.93  & 87.12 & \textbf{88.57}\\
\bottomrule
\end{tabular}
}
\label{tab6}
\end{table}

\noindent\textbf{ROI-level glomerular detection and segmentation.} 
The GS dataset, a paired human kidney slice dataset with four virtual stain registrations and manually annotated glomerular masks, was utilized to validate the ROI-level performance of VPGAN. Following the methodology of GramGAN, H\&E was virtually stained into MAS, PAS, and PASM, and glomerular detection and segmentation were implemented using Mask R-CNN~\cite{mask_rcnn}. As shown in Table 5, VPGAN-based virtual staining significantly enhances detection and segmentation performance, particularly in the H\&E2PASM task.

\noindent\textbf{Patch-level glomerular segmentation.} 
We performed patch-level glomerulus segmentation following KPIs, using four PAS-stained viral sub-datasets with distinct color variations, as shown in Figure \ref{fig6}(b). We hypothesized VPGAN normalization could improve segmentation, validated in Table \ref{tab6}. Average reflects mean test results from separate sub-dataset training, merge represents combined dataset training results, and normalization denotes pre-merge VPGAN data normalization, highlighting our method's potential.

\section{Conclusion}
In summary, we have introduced a novel unpaired slice virtual staining model designed for the virtual staining of pathological image slices. Our approach employs multiple Vision Language Model (VLM)-based prompts to achieve staining domain delineation and enhancement that aligns with actual pathological characteristics. It is crucial to highlight that we are the first in the pathological field to utilize contrastive learning methods to describe the complexity information in the staining process, thereby enhancing the staining effects. Additionally, we have provided a VLM-revised inference enhancement scheme to mitigate the risk of staining domain collapse. Our method has demonstrated the effectiveness of VLM-assisted virtual staining tasks and has been proven to serve as a data augmentation method for downstream tasks, such as glomerulus detection and segmentation. Importantly, the boundaries of VLM-based pathological prompt tuning tasks have been expanded by us, and more prompt schemes in virtual staining have been showcased.

\begin{acks}
This project was funded by the National Natural Science
Foundation of China 82090052.
\end{acks}

\bibliographystyle{ACM-Reference-Format}
\balance
\bibliography{reference}

\appendix
\section{Dataset Description}
\noindent\textbf{Stain Dataset.} 
 In the ANHIR dataset~\cite{anhir}, there are five sets of high-resolution tissue slides of the human kidney. Each set has four successive slides of the same tissue stained with different types of stain: H\&E, MAS, PAS, and PASM. We use four sets (Patient 1, Patient 2, Patient 3, and Patient 4) as the training set and one set (Patient 5) as the testing set. Following the settings in UMDST~\cite{UMDST} and DPI~\cite{DPI}, H\&E stained samples from Patient 1 were excluded from training due to large color and staining differences in the slices.  Each slide is cropped into a series of $256\times256$ patches with an overlap of 192, where the background regions (saturation<15) are discarded, and all remaining patches are used to train and test our model. There are 40,258 patches in the training set (7,688 for H\&E, 12,132 for MAS, 1,1458 for PAS, and 8,980 for PASM), and 8,070 patches in the testing set (1,989 for H\&E, 2,062 for MAS, 1,900 for PAS, and 2,119 for PASM). All methods are trained on the training set and tested on the test set. This also ensures that the epochs and iterations are the same, enabling a more fair comparison.

\noindent\textbf{Downstream Tasks Dataset.}  We tested the performance of our method on two glomerular detection and segmentation datasets, GS~\cite{GramGAN} and KPIs~\cite{kpis}, under different levels of images and various tasks. On the GS dataset, referring to the dataset settings in GramGAN, we divided the dataset into a training set and a test set at a ratio of 4:1. We randomly extracted patches of different sizes, ranging from 200×200 to 1,000×1,000, to test the performance of virtual staining in the object detection task at the ROI level. On the KPIs dataset, we adhered to the initial settings of the dataset. We achieved color difference normalization of PAS-stained mouse slices under four states: 56NX, DN, NEP25, and Normal, to improve the performance of detection and segmentation. 

\section{Baseline Detail}
\subsection{DeepSeek-R1} 
Based on comprehensive considerations of performance, cost efficiency, and other factors, we selected DeepSeek-R1~\cite{deepseekR1} as the large language model (LLM) for concept anchoring generation. We attribute the success of our approach to its several key advantages:

First, it demonstrates exceptional role-playing capabilities, effectively adopting the professional perspective of a pathologist in its text outputs. Second, its web search functionality not only addresses the prevalent hallucination issues in the medical domain but also enables a more contextually appropriate grasp of data and task-specific nuances. In fact, medical data, and even clinical scenarios, often exhibit strong regional specificity, shaped by complex factors such as geographical conditions, climate, and sociocultural environments. The web search feature allows the model to approximate the localized expertise of physicians. Finally, we must commend the model’s robust reasoning capabilities, which empower DeepSeek-R1 to deliver more comprehensive descriptions of stains or pathologies, encompassing a wealth of nuanced details.

\subsection{CycleGAN} 
The following constraints are satisfied by two mappings, $G_{AB}:A\to B$ and $G_{BA}:B\to A$, parameterized by neural networks, which are used by the CycleGAN model~\cite{CycleGAN} to estimate these conditionals. First, the output of each mapping
should match the empirical distribution of the target domain, when marginalized over the source domain. Then, mapping an element from one domain to the other, and then back, should produce a sample close to the original element. The former technique serves as the cornerstone of all generative adversarial networks (GAN)~\cite{GAN}. Mappings $G_{AB}$ and $G_{BA}$ are given by neural networks trained to fool adversarial discriminators $D_{B}$ and $D_{A}$, respectively. Enforcing marginal matching on target domain $B$, marginalized over source domain $A$, involves minimizing an adversarial objective with respect to $G_{AB}$:
\begin{equation}
\small
\mathcal{L}_{GAN}^{B}(G_{AB}, D_{B}) = \underset{b \sim p_{d}(b)}{\mathbb{E}} \left[ \log D_{B}(b) \right]+\underset{a \sim p_{d}(a)}{\mathbb{E}} \left[ \log (1 - D_{B}(G_{AB}(a))) \right],
\tag{19}
\end{equation}
while the discriminator $D_B$ is trained to maximize it. A similar adversarial loss $\mathcal{L}_{GAN}^{A}(G_{BA}, D_{A})$ is defined for marginal matching in the reverse direction.

Cycle-consistency enforces that, when starting from a sample a from A, the reconstruction $a^{\star }$ = $G_{BA}(G_{AB}(a))$ remains close to the original a. For image domains, closeness between a and $a^{\star }$ is typically measured with $L_1$ or $L_2$ norms. When using the $L_1$ norm, cycle-consistency starting from A can be formulated as:
\begin{equation}
\small
\mathcal{L}_{CYC}^{A}(G_{AB}, G_{BA}) = \underset{a \sim p_{d}(a)}{\mathbb{E}} \left\| G_{BA}(G_{AB}(a)) - a \right\|_1.
\tag{20}
\end{equation}
And similarly for cycle-consistency starting from B. The full CycleGAN objective is given by:
\begin{equation}
\small
\begin{split}
\mathcal{L}_{GAN}^{A}(G_{BA}, D_{A}) + \mathcal{L}_{GAN}^{B}(G_{AB}, D_{B}) + \\
\nu   \mathcal{L}_{CYC}^{A}(G_{AB}, G_{BA}) + \nu   \mathcal{L}_{CYC}^{B}(G_{AB}, G_{BA}),
\end{split}
\tag{21}
\end{equation}
where $\nu$ is a hyper-parameter that balances between marginal matching and cycle-consistency.

CycleGAN's success can be attributed to the complementary roles of marginal matching and cycle-consistency in its objective. Marginal matching enforces realism per domain.

\subsection{DDIM Inversion} 
DDIM~\cite{DDIM} utilizes an implicit non-Markovian process for sample generation, differing from DDPM~\cite{DDPM} which relies on a Markovian chain. This non-Markovian approach enables accelerated sampling through step-skipping in the reverse diffusion process. The core reverse-process equation of DDIM is expressed as:

\begin{equation}
\begin{aligned}
\small
\mathbf{x}_{t-1} = \sqrt{\alpha_{t-1}} \left( \frac{\mathbf{x}_t - \sqrt{1 - \alpha_t} \boldsymbol{\epsilon}_\theta(\mathbf{x}_t, t)}{\sqrt{\alpha_t}} \right) \\ + \sqrt{1 - \alpha_{t-1}} \boldsymbol{\epsilon}_\theta(\mathbf{x}_t, t),
\end{aligned}
\tag{22}
\end{equation}
$\boldsymbol{\epsilon}_\theta$ is the network predicting noise at each step.

The inversion process involves running the DDIM sampling process in reverse, which can be formulated as:
\begin{equation}
\begin{aligned}
\small
\mathbf{x}_{t+1} = \sqrt{\alpha_{t+1}} \left( \frac{\mathbf{x}_t - \sqrt{1 - \alpha_t} \boldsymbol{\epsilon}_\theta(\mathbf{x}_t, t)}{\sqrt{\alpha_t}} \right) \\ + \sqrt{1 - \alpha_{t+1}} \boldsymbol{\epsilon}_\theta(\mathbf{x}_t, t).
\end{aligned}
\tag{23}
\end{equation}

We denote $\epsilon^*_t$ as the groundtruth of prediction.To enhance the inversion process, consider the following modifications:

\begin{equation}
\small
\mathbf{x}_0 = \mathbf{x}^*_t + \left( \frac{1}{\alpha_t} - 1 \right) \sigma, \quad \sigma > 0,
\tag{24}
\end{equation}

\begin{equation}
\small
\boldsymbol{\epsilon}_\theta(\mathbf{x}_t, t) = \epsilon^*_\theta(\mathbf{x}_t, t) + \sigma, \quad \sigma > 0.
\tag{25}
\end{equation}

In non-conditional DDIM inversion, larger time steps help reduce error since error decreases as $t$ increases. The process works by iteratively applying the equations to trace the sample back to its original noise vector.

\section{Experimental parameter settings}
Most parameter settings have already been configured in the code. Please refer to the code content for details. To ensure fair comparison and demonstrate optimal performance, we follow all parameter settings of the baseline~\cite{CycleGAN,DPI}. Some hyperparameters in our method may have different numerical values depending on the dataset, as shown in the table \ref{tab7} below.

\begin{table}[h]
\centering
\caption{The hyperparameter values of our method on various datasets. }
\resizebox{0.45\textwidth}{!}{%
\begin{tabular}{c|ccc}
\toprule
hyperparameter setting & H\&E2MAS & H\&E2PAS & H\&E2PASM  \\
\midrule
$\alpha$ & 30 & 50 & 30 \\
$\beta$ & 0.1 & 0.1 & 0.1 \\
$\gamma$ & 0.1 & 0.1 & 0.1 \\
$\mu$ & 0.05 & 0.55 & 0.8 \\
$\lambda$ & 0.001 & 0.001 & 0.05 \\

\bottomrule

\end{tabular}
}
\label{tab7}
\end{table}

\section{Limitation and Future Work}
\subsection{Limitation} 
\noindent\textbf{Inference Time Burden.}  Although our inference enhancement achieves strong performance, DDIM still incurs substantial computational overhead even with just 50 sampling steps, requiring 5–10 minutes to enhance a single 256×256 image. This imposes a prohibitive time burden for whole-slide image (WSI) staining enhancement. We experimented with several acceleration methods for diffusion models, but all led to degraded inference quality.  

\noindent\textbf{ResNet CLIP not Specialized for Pathology.}  Our stain normalization method leverages multi-level MSE based on ResNet101~\cite{resnet} CLIP~\cite{CLIP}, an approach that is difficult to replicate with ViT-based architectures like CLIP, CONCH~\cite{CONCH}, or MUSK~\cite{MUSK}. We attribute this advantage to ResNet’s hierarchical structure, which effectively captures diverse visual features across different scales. However, a notable limitation is that all current pathology-specific vision-language model (VLM) adopt ViT architectures. This prevents direct comparison between natural image optimized ResNet CLIP and pathology specialized ResNet encoders, where the former excels in color perception while the latter better understands histopathology.

\subsection{Future Work}
We aim to explore the greater potential of VLMs and Diffusion models in pathological downstream tasks, extending to more diverse staining or other tasks, while seeking tighter integration methods between the two technologies. 

On the other hand, we also look forward to implementing more diverse multimodal data fusion approaches, where images, text, bulk RNA-seq data, spatial transcriptomics, and other information can be synergistically integrated to achieve superior results.

\end{document}